\documentstyle[aps,preprint,psfig]{revtex}
 
\begin{document}
 
\title{Stable integration of
isolated cell membrane patches in a nanomachined aperture: a step towards
a novel device for membrane physiology}

\author{N. Fertig, A. Tilke, R. H. Blick, and J. P. Kotthaus}
\address{Center for NanoScience and Sektion Physik,
Ludwig-Maximilians-Universit\"{a}t,
Geschwister-Scholl-Platz~1\\80539 Munich, Germany.}
\author{J. C. Behrends, and G. ten Bruggencate}
\address{Physiologisches Institut, Ludwig-Maximilians-Universit\"{a}t,\\
Pettenkoferstra\ss e~12,\\
80336 Munich, Germany.}

\date{\today}
\maketitle

\begin{abstract}
We investigate the microscopic contact of a cell/semiconductor hybrid. 
The semiconductor is nanostructured with the aim of single channel 
recording of ion channels in cell membranes. This approach will overcome
many limitations of the classical patch-clamp technique. The integration of silicon-based 
devices 'on-chip' promises novel types of experiments on single ion channels.
\end{abstract}


\newpage

Nanostructuring allows to build devices with dimensions similar to 
those of basic biological units, e.g. ion channels in cell membranes.
Ion channels are proteins that are integral parts of cell membranes 
and act as pores. They regulate the flow of ions in- and out of the 
cell \cite{hille 92}. Exhibiting different kinds of gating mechanisms 
they act as basic excitable units in biological systems.
The function of these elementary units is therefore of fundamental 
importance for information processing in neural systems.\\

For about two decades physiologists have been able to resolve ionic 
currents through single ion channels by using the patch-clamp technique 
\cite{neher 76,hamill 81}. This method relies on forming a $\mu$-size 
contact with the cell membrane by means of an electrolyte-filled glass 
pipette. The open tip of the pipette is pressed against the membrane, 
defining an isolated patch. Due to the strong glass-membrane adhesion 
\cite{opsahl 94}, a G$\Omega$-seal is obtained which allows current 
measurements with resolution of a few 100~fA. A basic limitation of 
this approach is the limited recording bandwith B$<100~$kHz. This 
limitation arises mainly because of stray capacitances and the high 
access resistance of the long-tapered pipette \cite{benndorf 95}.
In contrast, the geometry of the semiconductor based probes used in our
 approach should overcome these limitations. We define a nanoscale aperture 
located in a suspended Si$_3$N$_4$ membrane on micromachined silicon substrate. 
This enables us to minimize the distance between the ion channel under
 investigation and the recording electrode. In addition, with semiconductor 
structuring techniques, the passive glass pipette can be replaced by a versatile 
probe, which can easily integrate active semiconductor elements e.g. amplifiers 
or electromechanical devices. Finally, due to the open geometry of the probe, 
imaging techniques such as fluorescence microscopy, atomic force microscopy 
(AFM) and scanning electron microscopy (SEM) can be applied.\\
In Fig.~1(a) an SEM-micrograph of our device with an integrated cell 
membrane is depicted.
A Si$_3$N$_4$-layer is suspended on a micron scale by etching a V-groove 
in the (100)-silicon substrate beneath.
In order to build these suspended membranes we deposit a 120~nm thick 
Si$_3$N$_4$-layer on both sides of a (100) low n-doped silicon substrate 
using Low Pressure Chemical Vapor Deposition (LPCVD).
Applying standard optical lithography and Reactive Ion Etching (RIE) we 
define an etch mask on the backside of the samples. Subsequent anisotropic 
wet etching in a KOH-solution results in a V-shaped groove, where the upper 
Si$_3$N$_4$-layer serves as an etch stop. Adjusting the size of the etch 
mask we build a suspended Si$_3$N$_4$-layer with dimensions of a few ten 
microns side length. Both optical lithography as well as low-energy 
electron-beam lithography is used to define an orifice in the suspended 
membrane. The lithographic pattern is transferred into the membrane by 
an RIE process. The lower inset in Fig.~1(a) shows apertures in such a 
suspended membrane with sizes ranging from 500~nm down to 50~nm. Due to 
this nanostructuring process the geometry of the aperture can be freely chosen.\\
The integration of a cell membrane is achieved by positioning a cell on 
top of the probe. The device is installed into a classical patch-clamp 
setup including a remotely controlled positioning system and a microscope. 
The inset in Fig.~1(a) shows the schematical arrangement of the semiconductor-cell 
hybrid.
In order to carry out electrical measurements, the ensemble is connected 
via electrodes in standard Ringer's electrolyte solution (270 mOsm) 
forming the extra-cellular medium.
Cultured embryonic cells from rat striatum or C6-glioma cells are 
acutely dissociated applying standard trypsin treatment and trituration. 
A glass suction pipette is used to move an isolated cell onto the aperture
 as shown in Fig.~1(b). By applying negative pressure from below, the 
cell's membrane is partially sucked into the opening. This procedure is 
in close analogy to the standard patch-clamp technique. In order to 
obtain a cell-free patch, the glass pipette is used to remove the cell 
body, leaving an excised membrane patch in the aperture.
In Fig.~2, this excised patch is shown from both sides of the device. 
The micrographs were taken with a low-voltage scanning electron microscope 
(SEM) at resolution of about 1~nm. Fig.~2(a) shows a top view of the aperture. 
Cellular material (presumably cytoskeletal elements) is seen to fill the entire lumen.
Fig.~2(b) shows a close-up view of the cell membrane that has been dragged 
into the opening.
Imaging cellular structures is only possible after fixatation with 
glutaraldehyde solution, dehydration in graded alcohol and drying in 
a critical point drier. This procedure is responsible for the somewhat 
distorted surface structure of the cell membrane.
However, the image clearly shows, that there is an extremely close 
association of the membrane with the silicon nitride material without 
any visible gaps. This finding justifies the expectation that Si$_3$N$_4$
 can, in our design, substitute for glass in creating G$\Omega$-seals. It 
is also in line with the glass-like properties of this material. These 
adhesion properties are of great importance when interfacing neurons and 
silicon \cite{weis 96}. Single channel recording is only made possible by 
the so-called G$\Omega$-seal where the membrane sticks tightly to the glass 
of the pipette \cite{opsahl 94}. Demonstrating a G$\Omega$-seal with the Si$_3$N$_4$
membrane is therefore the next major step towards patch clamp recording with the 
device presented here.
Another advantage of our approach becomes obvious: due to its geometry our 
device lends itself 
to visualization techniques such as SEM, atomic force 
microscopy (AFM) or scanning nearfield optical microscopy (SNOM).\\

Furthermore, we applied confocal fluorescence microscopy for imaging the 
hybrid in the ionic solution i.e. in a sitation where the membrane proteins 
and their functions are intact. In order to visualize the membrane, 
we incubated isolated cells with a solution containing the fluorescent 
marker bis-oxanol prior to integrating the membrane into the probe. The 
fluorophore is excited by blue light (488~nm) emitted from a Ar-ion laser. 
In Fig.~3(a) a scanning micrograph of the membrane-semiconductor hybrid 
taken with a confocal fluorescence microscopy is shown. On the suspended 
Si$_{3}$N$_4$-layer fluorescent cell debris is found in the environment 
of the aperture. A structure of more regular, round shape can be discerned 
near the center of the image and represents fluorescent cellular membrane
 incorporated in the aperture. As shown in Fig.~3(b), using a z-scan 
series, this structure can be definitely distinguished from the surrounding 
debris: The graph shows a plot of the fluorescence intensity as a function 
of the distance of the confocal plane from the probe surface. Thus, successive 
optical sections parallel to the probe surface ranging from about $-4~\mu$m to 
$4~\mu$m with zero set at the Si$_{3}$N$_4$-membrane level are taken. The three
 curves correspond to the normalized fluorescence light intensity emitted from 
the clean Si$_{3}$N$_4$-film, the debris on top of the film and the membrane in 
the aperture, respectively. Obviously, the fluorescence of the fractured cell 
material is emitted starting from a $z$-position higher than that of the 
Si$_{3}$N$_4$-film. In contrast, the fluorescence of the incorporated membrane 
displays a $z$-range on the same level or even lower than the reference Si$_{3}$N$_4$-film.
 The increase of fluorescence intensity of the debris in the negative 
z-range is related to the backscattering of excitation light from the Si$_{3}$N$_4$-layer 
acting as a bifringent mirror. Since the integrated cell-membrane is freely suspended, 
this effect is not seen in the aperture.\\
In conclusion, we have shown a first realization of a cell membrane patch integrated 
into a nanostructured semiconductor device verified by fluorescence and SEM-micrographs. 
Attaching native cell membranes to nanostructured probes is the first step towards patch 
clamp recording with semiconductor or silicon-on-insulator (SOI) devices. In addition, 
the geometry of our hybrid enables various methods of microscopy such as confocal 
fluorescence or atomic force microscopy to be applied in situ. The application of 
the device presented for patch clamp recording will be discussed elsewhere 
\cite{niels 2010}. Furthermore, it has to be noted, that the principles of 
processing such a nano patch clamp (NPC) chip can easily be transferred to 
Silicon-on-Quartz or other material classes.
Combining the patch-clamp technique with semiconductor devices allows the 
integration of active amplifying devices, e.g. field effect transistors. By 
using lithographic methods, these active elements can be positioned in the 
immediate vicinity of the ion channel. The noise level of the measurement can 
thus be lowered dramatically due to an 'on chip' amplification, leading to a 
highly improved resolution.\\ \\
We would like to thank F. Rucker, E. Rumpel, L. Pescini, H. Lorenz, and H. Gaub for 
valuable discussions and support. Thanks are due to A. Kriele, G. Zitzelsberger 
A. Gr\"{u}newald and L. Kargl for expert techical support. 
The semiconductor wafers were kindly supplied by I. Eisele of the 
Universit\"{a}t der Bundeswehr (Munich). The nitride films were grown by H. Geiger 
of the Universit\"{a}t der Bundeswehr (Munich). This work was supported on part by 
the Deutsche Forschungsgemeinschaft (SFB 531).

\newpage

Fig. 1: (a) SEM micrograph of V-groove in (100)-silicon with 
a suspended Si$_3$N$_4$-layer on top. In the suspended Si$_3$N$_4$-layer a 
small aperture is nanostructured by optical or electron-beam-lithography and RIE. 
In the aperture cell material is incorporated. The upper inset depicts the 
schematical arrangement of the semiconductor-cell hybrid. The lower inset shows 
a series of holes in a suspended membrane with dimensions down to 50~nm.\\
(b) Photograph of the probe with cell positioned on top of the aperture. 
The glass pipette on the right is used to manipulate the cell.\\

Fig. 2: (a) Top-view of the aperture with incorporated cell material. The 
arrows indicate the circumference of the opening.\\ 
(b) Close-up of the cell membrane taken from the backside protruding 
from the opening. The membrane is sealed tightly to the Si$_3$N$_4$-layer 
with no remaining cleft in between.\\

Fig. 3: (a) Fluorescence scanning micrograph taken with a confocal microscope. 
The cell-membrane is labeled with fluorescent marker (thick arrow). Some 
cell debris is also visible (thin arrows).\\
(b) $Z$-series taken from the fluorescence of the cell-membrane, the 
Si$_3$N$_4$-layer and the debris on top of it (for details see text).\\

\end{document}